\documentclass[12pt]{amsart}
\usepackage{geometry} % see geometry.pdf on how to lay out the page. There's lots.
\usepackage{moreverb}
\usepackage[dvips,colorlinks,bookmarksopen,bookmarksnumbered,citecolor=red,urlcolor=red]{hyperref}
\usepackage{multirow}
\usepackage{graphicx}

\newcommand{\bA}{\mathbf{A}}
\newcommand{\bC}{\mathbf{C}}
\newcommand{\bM}{\mathbf{M}}
\newcommand{\bK}{\mathbf{K}}
\newcommand{\bphi}{\boldsymbol{\phi}}

\title[Rayleigh damping in inelastic analyses]{Initial vs. tangent stiffness-based Rayleigh damping in inelastic time history seismic analyses}

\date{} % delete this line to display the current date

\begin{document}

\maketitle

\begin{center}
P.~Jehel\footnote{Laboratoire MSSMat /CNRS-UMR 8579, \'Ecole Centrale Paris, Grande voie des Vignes, 92295 Ch\^atenay-Malabry Cedex, France}\footnote{Department of Civil Engineering and Engineering Mechanics, Columbia University, 630 SW Mudd, 500 West 120th Street, New York, NY, 10027, USA}, P.~L\'eger\footnote{Department of Civil Engineering, \'Ecole Polytechnique de Montr\'eal, P.O. Box 6079, Station CV, Montreal, QC, H3C 3A7, Canada} and A.~Ibrahimbegovic\footnote{LMT-Cachan (ENS Cachan/CNRS/UPMC/PRES UniverSud Paris), 61 avenue du Pr\'esident Wilson, 94235 Cachan Cedex, France} \\
\vspace{0.5cm}
pierre.jehel[at]ecp.fr \\
\vspace{0.5cm}
\small{Paper accepted for publication in \emph{Earthquake Engineering and Structural Dynamics}} \\
\small{Published online (wileyonlinelibrary.com). DOI: 10.1002/eqe.2357}
\end{center}

%%%%%%%%%%%%%%%%%%%%%%%%%%%%%%%%%%%%%
%\begin{document}

%\address{
%\affilnum{1}Laboratoire MSSMat /CNRS-UMR 8579, \'Ecole Centrale Paris, Grande voie des Vignes, 92295 Ch\^atenay-Malabry Cedex, France\break
%\affilnum{2}Department of Civil Engineering and Engineering Mechanics, Columbia University, 630 SW Mudd, 500 West 120th Street, New York, NY, 10027, USA\break
%\affilnum{3}Department of Civil Engineering, \'Ecole Polytechnique de Montr\'eal, P.O. Box 6079, Station CV, Montreal, QC, H3C 3A7, Canada\break
%\affilnum{4}LMT-Cachan (ENS Cachan/CNRS/UPMC/PRES UniverSud Paris), 61 avenue du Pr\'esident Wilson, 94235 Cachan Cedex, France
%}
%
%\corraddr{pierre.jehel[at]ecp.fr}

%%%%%%%%%%%%%%%%%%%%%%%%%%%%%%%%%%%%%
\begin{abstract}
In the inelastic time history analyses of structures in seismic motion, part of the seismic energy that is imparted to the structure is absorbed by the inelastic structural model, and Rayleigh damping is commonly used in practice as an additional energy dissipation source. It has been acknowledged that Rayleigh damping models lack physical consistency and that, in turn, it must be carefully used to avoid encountering unintended consequences as the appearance of artificial damping. There are concerns raised by the mass proportional part of Rayleigh damping, but they are not considered in this paper. As far as the stiffness proportional part of Rayleigh damping is concerned, either the initial structural stiffness or the updated tangent stiffness can be used. The objective of this paper is to provide a comprehensive comparison of these two types of Rayleigh damping models so that a practitioner i) can objectively choose the type of Rayleigh damping model that best fits her/his needs and ii) is provided with useful analytical tools to design Rayleigh damping model with good control on the damping ratios throughout inelastic analysis. To that end, a review of the literature dedicated to Rayleigh damping within these last two decades is first presented; then, practical tools to control the modal damping ratios throughout the time history analysis are developed; a simple example is finally used to illustrate the differences resulting from the use of either initial or tangent stiffness-based Rayleigh damping model.
\end{abstract}

\textbf{keywords:} Rayleigh damping, inelastic structure, modal analysis, damping ratio time history, upper and lower bounds for damping ratios.

%%%%%%%%%%%%%%%%%%%%%%%%%%%%%%%%%%%%

\section{Introduction}

Seismic analyses were first developed for elastic structures. An elastic structure does not absorb energy and, therefore, damping was added to represent all the energy dissipation sources at the boundary conditions or at small scales not explicitly considered for civil engineering applications (a list of such energy dissipating phenomena can be found in~\cite{NasJonHen1985}). For this purpose, Rayleigh damping is mathematically very convenient because, once projected onto the undamped modal basis, the set of equations of motion of the discrete structures are uncoupled. This type of damping is herein referred to as \emph{elastic damping}. Then, to account for yielding mechanisms in the overall structural response, elastic viscous-equivalent seismic analyses appeared~\cite[p.74]{CloPen1975}. In this case, the structural model still is elastic but the damping properties are enhanced so as to account for both elastic damping and energy dissipation due to the actual, but not explicitly modeled, yielding response. This latter type of damping is herein referred to as \emph{hysteretic damping}.

However, inelastic time history analysis (ITHA), which are based on the building of an inelastic structural model, is the most appropriate way to properly account for hysteretic mechanisms in seismic analyses. In this case, the structural modes are not uncoupled anymore in the modal basis. The modal basis could be updated after each inelastic event to maintain uncoupled modes, but the computational benefit would be counterbalanced by the additional cost resulting from the successive modal analyses required. Consequently, there is few mathematical advantage in using Rayleigh damping for ITHA (the main advantage is that there is no need to explicitly build and store a damping matrix because mass and stiffness matrices already are stored for other purposes). In spite of this, Rayleigh damping still is commonly used in ITHAs. Ideally, Rayleigh damping should be added to model elastic damping only, and hysteretic damping should arise from the explicit modeling of the energy dissipation mechanisms in the inelastic structural model. In practice, this can hardly be achieved because, on the one hand, controlling the amount of elastic damping in ITHA is challenging due to the intrinsic nature of Rayleigh damping models and, on the other hand, inelastic structural models only provide an approximation of the numerous inelastic phenomena that actually contribute to the seismic energy absorption in the structure. This paper only focusses on the issue of maintaining good control on elastic Rayleigh damping throughout inelastic analysis.

In its most general form, Rayleigh damping consists in adding viscous forces of the form $\mathbf{f}^D(t) = \mathbf{C}(t) \dot{\mathbf{u}}(t)$ in the discrete structural equations of motion, where $\dot{\mathbf{u}}$ is the displacements rate and the damping matrix $\mathbf{C}(t)$ is built as a linear combination of the structural mass and stiffness matrices $\mathbf{M}$ and $\mathbf{K}$:
\begin{equation} \label{eq:Ray-dam-mat}
 \mathbf{C}(t) = \alpha(t) \mathbf{M} + \beta(t) \mathbf{K}(t) \ .
\end{equation}
The Rayleigh coefficients $\alpha$ and $\beta$ are computed so that the critical damping ratios $\xi_A$ and $\xi_B$ are observed at the frequencies $\omega_A$ and $\omega_B$. $\alpha$ and $\beta$ can be either set once and then frozen or updated throughout ITHA. The stiffness matrix is commonly built either from the initial or from the updated tangent stiffnesses. Both mass- and stiffness-proportional terms of Rayleigh damping models can generate difficulties in controlling elastic damping throughout an inelastic analysis. Contrary to the stiffness-proportional term, which is difficult to control in inelastic analyses only, problems can arise from the mass-proportional component no matter whether it is an elastic(-equivalent) or an inelastic time-history analysis. Indeed, assuming that the mass matrix is diagonal, the physical counterpart of a mass-proportional term is a viscous damper connecting a structural degree of freedom to the frame which contains the reference point for measuring displacements in the structure. Then, the mass-proportional term can lead to unrealistically high damping forces whenever the whole or parts of the structure behave like a rigid body. This issue encountered with the mass-proportional term has already been well explained and illustrated~\cite{Hall2006}. Consequently, in this paper, although both mass- and stiffness-proportional terms are used to construct the damping matrix, only the stiffness-proportional component is focussed on. 

How to avoid spurious damping forces and improve control on damping throughout inelastic analyses has been widely studied and led to practical recommendations (\emph{e.g.} \cite{Hall2006,LegDus1992,Charney2008}). However, when it comes to dealing with the stiffness-proportional term, one can find in the literature differing viewpoints on whether to use initial or tangent stiffness. The objective of this paper is to provide a comprehensive comparison of Rayleigh damping based on either initial or tangent stiffness, and to answer the question: ``which of initial and tangent stiffness-based Rayleigh damping provides better control on damping ratio throughout ITHA?" (the mass-proportional term being also present in the Rayleigh damping models). To that aim, we first present in the next section a review of the literature where the different strategies that have been proposed to build the Rayleigh damping stiffness-proportional term are exposed. Then, in section 3, mathematical developments lead to formulas that allow quantifying damping ratios shifts due to stiffness degradations. These relations are useful to design a Rayleigh damping model before running an analysis and, once the analysis has been run, to assess the validity of the elastic damping modeling. To the best of our knowledge, this is the first time such formulas are provided for initial stiffness-based Rayleigh damping. For the sake of completeness, we already mention here that the formula we derive in section 3 are only valid when the two following assumptions are made: i) the damping ratios $\xi_A$ and $\xi_B$ chosen to identify the Rayleigh coefficients $\alpha$ and $\beta$ are taken as equal ($\xi_A = \xi_B$), and ii) the structural equations of motion are uncoupled when expressed in modal coordinates, which is in most of the cases only an approximation for inelastic structures with initial stiffness-based Rayleigh damping. Those two assumptions are common for the type of problems we consider in this work and they do not reduce the contributions of this paper comparing to previous publications on the subject. Before closing the paper with some conclusions, we compare the performances of initial and tangent stiffness-based Rayleigh damping models in the analysis of a simple structure with stiffness degradations.

%%%%%%%%%%%%%%%%%%%%%%%%%%%%%%%%%%%%%
\section{In the literature}

According to Charney~\cite{Charney2008}, one can go back to the eighties to find the first papers dedicated to problems pertaining to modeling damping in inelastic structures with the work of Chrisp in 1980 \cite{Chrisp1980} and Shing and Mahin in 1987 \cite{ShiMah1987}.

In the nineties, L\'eger and Dussault~\cite{LegDus1992} investigate, in the case of inelastic frame structures in seismic loading, the variation of nonlinear response indicators (average ductility, hysteretic-to-input energy ratio, average number of yield incursions) according to the additional damping model used. A bilinear hysteresis model is assigned to the structural elements. Amongst other mass- or stiffness-proportional damping models, Rayleigh damping models either with elastic or tangent stiffness and either with frozen or updated coefficients $\left( \alpha, \beta \right)$ are considered. The authors have developed a computer program to update $\alpha$ and $\beta$ at each time step. This allows maintaining constant critical damping ratio throughout the seismic analyses for the two frequencies used to identify the Rayleigh coefficients, which are, in this work, the first natural frequency and the frequency for which 95\% of effective modal mass is represented by the  truncated eigenbasis. Note that these frequencies are not the initial elastic ones but that they also are updated at each time step. L\'eger and Dussault show that, while having little effects on the amount of energy imparted to a structure by an earthquake, the choice of an additional damping model significantly influences the amount of hysteretic energy due to damage in the structure. The term ``additional damping'' is used here to refer to a source of damping that comes in addition to the damping resulting from the absorption of hysteretic energy in the structural model. L\'eger and Dussault use for instance the displacement ductility averaged over all the stories as an indicator of the inelastic structural response and, for the El Centro ground motion considered, variations of this latter indicator of up to 40\% from one damping model to the other are observed. Considering the Rayleigh model with updated coefficients as a baseline, recommendations on which additional damping model to use given the elastic fundamental period of the structure close the paper.

From the statement that the number of degrees of freedom (DOFs) needed to assemble the stiffness matrix is often much larger than that needed for building an adequate mass matrix, Bernal~\cite{Bernal1994} shows that spurious damping forces are likely to arise from the presence of massless -- or with relatively small inertia -- DOFs in inelastic structural systems. Indeed, massless DOFs have the tendency to undergo abrupt changes in velocity when stiffness changes, leading to unrealistically large viscous damping forces. In this work, Bernal adds proportional damping in the equilibrium equations using the Caughey series. It is stated that the most important effect of spurious damping forces is found in the distortion they introduce in the maximum internal forces rather than the displacements. As a solution to prevent spurious damping due to massless DOFs, it is suggested to assemble the damping matrix using the stiffness matrix condensed to the size of the DOFs with mass, followed by expansion to the full set of coordinates with columns and rows of zeros. This procedure is equivalent to what is proposed in other works where it is expressed in the other terms: assembling the damping matrix, element by element, with zero damping assigned to the DOFs where local abrupt changes of stiffness can occur (see \emph{e.g.} the procedure advocated in~\cite{Perform3D} for structures with fiber elements, or the issues raised by local stiffness changes in~\cite{Charney2008}).

In \cite{Hall2006}, Hall focuses on practical situations where the use of Rayleigh damping can lead to damping forces that are unrealistically large compared to the restoring force, and then proposes a capped viscous damping formulation to overcome some of the problems pointed out. It is stated in this paper that the tangent stiffness should not be used to build the Rayleigh damping matrix, especially because of the convergence difficulties it can cause. To illustrate this assertion, the author gives the example of the local damping stress, resulting from the stiffness-proportional part of Rayleigh damping, that can jump from zero to a possibly large value in case of crack closing as soon as contact is made. Then, Hall distinguishes between problems arising from the mass- and stiffness-proportional part of Rayleigh damping and quantifies the likely undesirable effects. The mass-proportional term can lead to unrealistically high damping forces whenever the whole or parts of the structure behave like a rigid body: formulation of an earthquake analysis in terms of total motion, superstructure on a relatively flexible base, portions of a structure that break loose like in a dam undergoing sliding at its base,... As far as the stiffness-proportional term of Rayleigh damping is concerned, very large damping forces can be generated when Rayleigh damping is based on the initial stiffness, while structural elements yield, leading to an increase in the velocity gradient: gravity dam undergoing cracking, presence of penalty elements,... Finally, as a remedy to the problems listed in the paper, the author proposes a capped viscous damping formulation in which the mass-proportional contribution is eliminated and the stiffness-proportional contribution is limited by bounds defined in accordance with the actual physical mechanism that limits the structural restoring forces.

In~\cite{Charney2008}, Charney first investigates the effects of global stiffness changes on the seismic response of a 5-story structure when Rayleigh damping is used. Two cases are considered: a reduction of the story stiffness by the same factor for each story, which does not change the mode shapes, and a nonuniform story stiffness reduction along the height, which leads to mode shapes that are different from those of the initial structure. In both cases, it is shown that, when structure yields, i) artificial damping is generated when the Rayleigh damping matrix is computed according to the initial structural properties; ii) significant but reduced artificial damping is generated when Rayleigh damping is built with tangent stiffness and fixed Rayleigh coefficients computed from the initial stiffness; iii) virtually no artificial damping is generated when Rayleigh damping is based on both tangent stiffness and updated coefficients from the tangent stiffness.

Then, Charney~\cite{Charney2008} also investigates the effects of local stiffness changes on the seismic structural time-history response when initial stiffness-based Rayleigh damping is used. The same 5-story structure as previously is considered, but this time with inelastic rotational springs at beam-to-column connections, which represent local yielding mechanisms. There is no rotational mass and zero hardening after yielding. Under these conditions, three different pairs of beams/columns stiffness and connections elastic stiffness are defined so that, in each case, the overall stiffness matrix of the structure be unchanged. The response histories are identical when a linear analysis is performed and when the damping matrix is built so that the rows and columns pertaining to the rotational DOFs are filled with zeros. However, when the Rayleigh damping matrix is built with non-zero initial stiffness associated to the massless DOFs, artificial viscous damping forces develop. These forces can be extremely high, especially as the initial spring stiffness is large. Yet they are not easy to detect.

To Charney, the best strategy to model damping would be of course to eliminate the use of viscous damping, because it is not physically sound, and replace it by frictional or hysteretic devices. Nevertheless, amongst other recommendations, Charney advocates that Rayleigh damping based on tangent stiffness be used. Indeed, with this method, the problems associated with local stiffness changes are always eliminated, and reduced artificial damping is generated because of global stiffness changes, which, moreover, can still be reduced by anticipating the shift of the structural natural frequencies. If elastic stiffness-based Rayleigh damping only is implemented in the computer program used, the damping matrix should be computed with zero stiffness associated to the elements that have large initial stiffness and that are likely to yield.

The more recent papers which also focus on modeling structural damping in inelastic time history seismic analyses are those of Zareian and Medina~\cite{ZarMed2010} and~Erduran \cite{Erduran2012}. In these papers, practical strategies are presented to cope with the problems encountered with Rayleigh damping in the context of performance-based seismic design. In the approach presented in~\cite{ZarMed2010}, each structural element is modeled with an equivalent combination of an elastic element with initial stiffness-proportional damping and yielding springs at the two ends without stiffness-proportional damping. With this strategy, i) numerical solution instabilities when significant changes in stiffness values occur are avoided because initial stiffness matrix is used in the damping model, and, also, no local spurious damping forces are generated in the structural parts that yield because there is no stiffness-proportional part in the damping model pertaining to these parts.

In~\cite{Erduran2012}, Erduran recalls that it has been shown that designing Rayleigh damping models based on the initial stiffness matrix results in unreasonably high damping forces after yielding and consequently decides to exclusively use tangent stiffness-based Rayleigh damping models in his study. This study consists in analyzing the story drift ratios, floor accelerations and damping forces in two 3- and 9-story steel moment-resisting frame buildings for three seismic hazard levels. It is concluded in~\cite{Erduran2012} that, as long as they are designed according to reduced modal frequencies (comparing to the elastic properties), Rayleigh damping models with both mass- and tangent stiffness-proportional components lead to reasonable damping forces and floor acceleration demands without suppressing higher modes effects in the 9-story building.

Either following or initiating these research efforts, there are computer program user manuals and technical reports who directly provide practitioners with advanced tools and guidelines for proper implementation of Rayleigh damping in inelastic time history seismic analyses. For instance: eight types of damping models are implemented in the computer program RUAUMOKO, whose user manual also comes with comprehensive discussion on damping \cite[pp. 9-10]{Ruaumoko}; visualization tools for damping effects in the simulations are available in the computer program PERFORM-3D, and an entire chapter is dedicated to modeling damping in the accompanying user guide \cite[\S 18]{Perform3D}; a large share is dedicated to viscous damping for inelastic seismic analyses in the technical reports~\cite[\S 2.4.2]{PEER-2010/111} and, to a more limited extent, in~\cite[\S 6.4.4]{FEMA-P695}.

From the review of the literature above, it is obvious that some authors recommend using initial stiffness-based Rayleigh damping models while others recommend using tangent stiffness. In the following section, we provide mathematical development for a rational comparison of the two approaches, relying on an analysis of the damping ratios time history throughout inelastic analyses.

%%%%%%%%%%%%%%%%%%%%%%%%%%%%%%%%%%%%
\section{Mathematical developments}

\subsection{Modal analysis for inelastic structures}

The equilibrium equations of a structure in seismic loading, with additional viscous damping, discretized in space, and transformed to undamped modal coordinates at time $t$, read:
\begin{equation} \label{eq:mod-equi}
 \boldsymbol{\Phi}^T_t \mathbf{M} \boldsymbol{\Phi}_t \ \ddot{\mathbf{U}}(t) + \boldsymbol{\Phi}^T_t \mathbf{C}(t) \boldsymbol{\Phi}_t \ \dot{\mathbf{U}}(t) + \boldsymbol{\Phi}^T_t \mathbf{K}(t) \boldsymbol{\Phi}_t \ \mathbf{U}(t) = \boldsymbol{\Phi}^T_t \left( \mathbf{F}^{sta}(t) + \mathbf{F}^{sei}(t) \right)
\end{equation}
where $\mathbf{U}(t)=\{U_m(t)\}_{m=1,..,N^m}$ is the undamped modal coordinates vector, $\mathbf{F}^{sta}(t)$ and $\mathbf{F}^{sei}(t)$ are the pseudo-static and seismic forces (the static force is applied step by step prior to the application of the seismic load to grasp possible nonlinear mechanisms), $\boldsymbol{\Phi}_t=\left( \phi_1(t) \ ... \ \phi_{N^m}(t) \right)$ is the matrix composed by the undamped modal shape vectors, and \tiny$\square$\normalsize$^T$ denotes the transpose operator. Mode shapes can also be computed from the damped system, but the undamped assumption is commonly retained for systems with low damping ratios, as it is the case in this work. Although cumbersome, the explicit dependence of every quantity on time $t$ is indicated (\tiny$\square$\normalsize$(t)$ or \tiny$\square$\normalsize$_t$) to emphasize that each of these quantities can change within the inelastic time history of the structure.

There are two ways to compute the viscous damping matrix. If \emph{modal damping} is used, the modal damping ratios $\xi_m$ are chosen at time $t$ when the modal analysis is performed, and the damping matrix is built as diagonal:
\begin{equation} \label{eq:mod-dam}
 \boldsymbol{\Phi}^T_t \mathbf{C}(t) \boldsymbol{\Phi}_t = \textrm{diag}(\mathcal{C}_m(t)) \quad \textrm{with} \quad \mathcal{C}_m(t) = 2 \ \xi_m(t) \omega_m(t) \mathcal{M}_m(t) \ .
\end{equation}
In this relation, we have introduced the undamped modal circular frequencies $\omega_m$ as $\omega_m^2(t) = \mathcal{K}_m(t) / \mathcal{M}_m(t)$, along with the notation $\mathcal{A}_m = \bphi_m^T \bA \bphi_m$ where $\bA = \bM$, $\bC$ or $\bK$. If \emph{Rayleigh damping} is used, a set of coefficients $\left( \alpha, \beta \right)$ is computed, and the damping matrix is built as:
\begin{equation} \label{eq:Ray-dam}
 \bC(t) = \alpha(t) \bM + \beta(t) \bK(t) \quad \Rightarrow \quad \mathcal{C}_m(t) = \alpha(t) \mathcal{M}_m(t) + \beta(t) \mathcal{K}_m(t) \ .
\end{equation}
From equations \eqref{eq:mod-dam} and \eqref{eq:Ray-dam}, we thus have the relation
\begin{equation} \label{eq:mod-Ray}
 \alpha(t) \bphi_m^T(t) \bM \bphi_m(t) + \beta(t) \bphi_m^T(t) \bK(t) \bphi_m(t) = 2 \ \xi_m(t) \omega_m(t) \bphi_m^T(t) \bM \bphi_m(t)
\end{equation}
between -- right-hand side -- the modal quantities that would arise from a modal analysis at any time $t$ in the history of the structure and -- left-hand side -- what actually results from the use of Rayleigh damping at this specific time $t$. Relation~\eqref{eq:mod-Ray} provides a definition of the modal damping ratio.

In equation~\eqref{eq:mod-equi}, both $\boldsymbol{\Phi}^T_t \mathbf{M} \boldsymbol{\Phi}_t$ and $\boldsymbol{\Phi}^T_t \mathbf{K}(t) \boldsymbol{\Phi}_t$ are diagonal matrices. $\boldsymbol{\Phi}^T_t \mathbf{C}(t) \boldsymbol{\Phi}_t$ however can be non-diagonal, which implies that~\eqref{eq:mod-equi} is a set of coupled equations. A damping model that is not diagonalized in the modal basis is qualified as \emph{non-classical} or \emph{non-proportional}.  As shown in relation~\eqref{eq:mod-dam}, modal damping inherently is classical. Rayleigh damping is however not necessarily classical: if initial stiffness is used, off-diagonal terms appear in $\boldsymbol{\Phi}^T_t \mathbf{C} \boldsymbol{\Phi}_t$ in most of the cases when structure yields. Relation~\eqref{eq:mod-Ray} provides exact values of $\xi_m(t)$ for \emph{classical} damping only; otherwise, not all the damping energy is included in the damping ratios. Consequently, in the presence of non-classical damping, the definition of $\xi_m$ in~\eqref{eq:mod-Ray} is only valid if the off-diagonal terms in $\boldsymbol{\Phi}^T_t \mathbf{C} \boldsymbol{\Phi}_t$ can be neglected. For systems with low damping ratios (a few percents of critical), as it is the case in this work, the assumption that the equations in~\eqref{eq:mod-equi} can approximated as decoupled in the presence of non-classical damping generally is, explicitly or implicitly, regarded as valid in the literature (\emph{e.g.} in~\cite{Hall2006,Charney2008}). Accordingly, we will hereafter retain this assumption.

\subsection{Implementing Rayleigh damping}

In practice, there are several methods to build a Rayleigh damping matrix, which all have different consequences on the time history of the damping ratios throughout the inelastic analysis:
\begin{enumerate}
 \item[$a$.] The Rayleigh coefficients, as well as the stiffness matrix, are set once for all before the beginning of the dynamic analysis. We refer to these quantities as $\left( \alpha_0, \beta_0 \right)$ and $\bK_0$.
These quantities are not necessarily identified considering the state of the structure at the beginning of the dynamic analysis: based on an approximation of the reduced stiffness, the state of the structure at any time within or at the end of the seismic loading can be used for the identification of the Rayleigh coefficients before running the dynamic analysis. At a given time $t$ in the structure time history, equation \eqref{eq:mod-Ray} thus takes the following expression:
\begin{equation}
 \xi^a_m(t) = \frac{1}{2 \ \omega_m(t)} \left( \alpha_0 + \beta_0 \ \frac{\bphi_m^T(t) \bK_0 \bphi_m(t)}{\mathcal{M}_m(t)} \right)
\end{equation}
which we rewrite as
\begin{equation} \label{eq:xi-a}
 \xi^a_m(t) = \frac{1}{2} \left( \frac{\alpha_0}{\omega_m(t)} + \beta_0 \ h_m^a(t) \omega_m(t) \right) \quad \textrm{with} \quad h^a_m(t) = \frac{\bphi_m^T(t) \bK_0 \bphi_m(t)}{\mathcal{K}_m(t)} \ .
\end{equation}
Note that $\boldsymbol{\Phi}^T_t \mathbf{K}_0 \boldsymbol{\Phi}_t$ generally is not diagonal and that the off-diagonal terms are neglected in this definition of $\xi_m^a(t)$.

 \item[$b$.] The Rayleigh coefficients are set once for all at the beginning of the dynamic analysis, as in case $a$, and the tangent stiffness matrix $\bK(t)$, updated at each time step, is used to build $\bC(t)$. With this definition, Rayleigh damping is classical throughout the analysis and we have:
\begin{equation} \label{eq:xi-b}
 \xi^b_m(t) = \frac{1}{2} \left( \frac{\alpha_0}{\omega_m(t)} + \beta_0 \ \omega_m(t) \right) \ .
\end{equation}

 \item[$c$.] Both the Rayleigh coefficients and the stiffness matrix are updated at each time step in the numerical analysis. In this case, the modal damping ratio $\xi^c_m(t)$ can be controlled throughout the analysis, \emph{e.g.} maintained constant by updating $\left( \alpha, \beta \right)$ accordingly:
\begin{equation} \label{eq:xi-c}
 \xi^c_m(t) = \frac{1}{2} \left( \frac{\alpha(t)}{\omega_m(t)} + \beta(t) \ \omega_m(t) \right) \ .
\end{equation}

 \item[$d$.] There is no risk of generating spurious local damping forces when methods based on the updated tangent stiffness are used to implement Rayleigh damping~\cite{Charney2008}. However, this risk arises when, as in case $a$, the Rayleigh coefficients along with the stiffness matrix are set once for all before the beginning of the dynamic analysis, and there are some elements in the structure with artificially large initial stiffness (plastic hinges, gap or contact elements). It is then advocated, \emph{e.g.} in~\cite{Charney2008}, to associate reduced $\beta_0$ factors to this latter type of element ($\beta_0^e < \beta_0$). However, this method can potentially generate the same type of global effects as method $a$. Indeed, defining a reduced initial stiffness matrix $\bK_0^r$ as:
 \begin{equation}
  \mathbf{C} = \mathsf{A}_e \left( \alpha_0 \mathbf{M}^e + \beta^e \mathbf{K}_0^e \right) = \alpha_0 \mathbf{M} + \beta_0 \mathbf{K}_0^r \quad \textrm{with} \quad \beta_0 \bK_0^r = \mathsf{A}_e \beta_0^e \bK_0^e \ ,
 \end{equation}
where $\mathsf{A}_e$ denotes the finite element assembly procedure \cite{Ibrahimbegovic2009}, it comes from equation~\eqref{eq:mod-Ray}:
\begin{equation} \label{eq:xi-d}
 \xi^d_m(t) = \frac{1}{2} \left( \frac{\alpha_0}{\omega_m(t)} + \beta_0 \ h_m^d(t) \omega_m(t) \right) \quad \textrm{with} \quad h^d_m(t) = \frac{\bphi_m^T(t) \bK_0^r \bphi_m(t)}{\mathcal{K}_m(t)} \ .
\end{equation}
If the reduced initial stiffness $\mathbf{K}_0^r$ better approximates the tangent stiffness matrix than the initial stiffness $\mathbf{K}_0$, the global effects with method $d$ are reduced comparing to method $a$. Also, using this method, off-diagonal terms can appear in the generalized damping matrix; those terms are neglected in this definition of $\xi_m^d(t)$.
\end{enumerate}

For methods $b$ and $c$, the updated stiffness matrix could loose its positiveness if significant softening or second-order effects would occur. Mathematically, it would imply that the modal frequencies be null or imaginary ($\omega^2 \leq 0$), which, physically, would not make any sense. However, this would only happen if the structure became unstable and, consequently, this is an ultimate potential problem that we let out of the scope of this work.

Method $c$ is rarely used in practice because a modal analysis has to be carried out after each inelastic event. Therefore, hereafter, we do not consider case $c$ anymore and the Rayleigh coefficients $\left( \alpha_0 , \beta_0 \right)$ are set once for all before the dynamic analysis.

\subsection{Rayleigh coefficients computation}

For method $i= a$, $b$ or $d$, we have, according to what precedes:
\begin{equation} \label{eq:xi-abcd}
 \xi^i_m(t) = \frac{1}{2} \left( \frac{\alpha_0}{\omega_m(t)} + \beta_0 \ h_m^i(t) \omega_m(t) \right) \quad \textrm{with} \quad h_m^b(t) = 1 \ , \ \forall m \ , \ \forall t \ .
\end{equation}
Two independent equations can be written from equation \eqref{eq:xi-abcd} with two different pairs of damping ratios and circular frequencies hereafter referred to as $(\xi_A, \omega_A)$ and $(\xi_B, \omega_B)$. The resulting set of two equations can then be solved to obtain $\alpha_0$ and $\beta_0$. For inelastic analyses, the practitioner can choose the pairs $(\xi_A, \omega_A)$ and $(\xi_B, \omega_B)$ either for two different modes $m_A$ and $m_B$ at the same time $t=t_A=t_B$, or for two different modes $m_A$ and $m_B$ at two different times $t_A$ and $t_B$, or for the same mode $m=m_A=m_B$ at two different times $t_A$ and $t_B$. $A$ and $B$ thus refer to a mode and a particular time in the history of the structure and, in the following, $A$ has to be understood as the pair $\left( t=t_A, m=m_A \right)$, and so for $B$.

For the sake of conciseness, we abandon the superscript $i$ that explicitly refers to method $a$, $b$ or $d$. Yet, one should keep in mind that the quantities $\xi_{A/B}$, $\omega_{A/B}$, $h_{A/B}$, $\alpha_0$ and $\beta_0$ all depend on the method used. Rewriting equation \eqref{eq:xi-abcd} for $A$ and $B$, we then have:
\begin{equation} \label{eq:xiA-xiB}
 \begin{Bmatrix}
  \xi_A \\
  \xi_B
 \end{Bmatrix} = \frac{1}{2}
 \begin{pmatrix}
  1 / \omega_A & h_A \ \omega_A \\
  1 / \omega_B & h_B \ \omega_B 
 \end{pmatrix}
 \begin{Bmatrix}
  \alpha_0 \\
  \beta_0
 \end{Bmatrix} \ .
\end{equation}

The Rayleigh coefficients can then be computed by inverting relation \eqref{eq:xiA-xiB}, which, as long as $h_A \omega_A^2 \neq h_B \omega_B^2$, is always possible:
\begin{equation} \label{eq:alp-bet}
 \begin{Bmatrix}
  \alpha_0 \\
  \beta_0
 \end{Bmatrix} = \frac{2 \ \omega_A \omega_B}{h_B \omega_B^2 - h_A \omega_A^2}
 \begin{pmatrix}
  h_B \omega_B & - h_A \omega_A \\
  - 1 / \omega_B & 1 / \omega_A
 \end{pmatrix}
 \begin{Bmatrix}
  \xi_A \\
  \xi_B
 \end{Bmatrix}.
\end{equation}
We assume that $\xi_A = \xi_B = \xi_0$, which will simplify the following developments. Because this assumption is commonly done in practice, it does not reduce the contributions of the present work. We then have:
\begin{equation}  \label{eq:alp-bet}
 \alpha_0 = \frac{2 \ \omega_A \omega_B \left( h_B \omega_B - h_A \omega_A \right)}{h_B \omega^2_B - h_A \omega^2_A} \ \xi_0 \qquad \textrm{and} \qquad \beta_0 = \frac{2 \left( \omega_B - \omega_A \right)}{h_B \omega^2_B - h_A \omega^2_A} \ \xi_0 \ .
\end{equation}

Note that the coefficients $h$ generally are absent in the definition of the Rayleigh coefficients. Common definitions for $\alpha_0$ and $\beta_0$ indeed are:
\begin{equation}
 \alpha_0 = \frac{2 \ \omega_A \omega_B}{\omega_A + \omega_B} \xi_0 \qquad \textrm{and} \qquad \beta_0 = \frac{2}{\omega_A + \omega_B} \xi_0 \ ,
\end{equation}
but this actually is only exact for case $b$ where $h^b_m(t)=1$ for all $t$ and $m$. There is another important notion that is generally not explicitly mentioned in the definition of the Rayleigh coefficients in ITHA and that we recall here: the index $A$ (respectively $B$) does not only refer to a specific mode but to a mode at a given time in the history of the structure.

\subsection{Upper and lower bounds for the damping ratios}

\begin{figure}[htb]
\centering
 \includegraphics[width=0.6\textwidth]{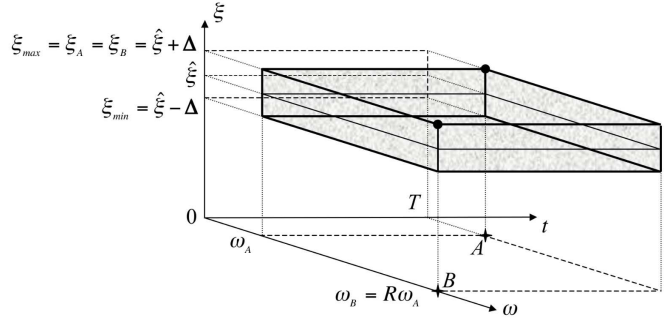}
\caption{We seek to characterize the grey box within which the damping ratios remain all along the inelastic analysis ($t \in [0 , T]$). $\hat{\xi}$ is the targeted damping ratio in the range $[\omega_A , \omega_B]$, $\Delta$ is the deviation of the actual damping ratio from the targeted one. The black points indicate two possible points to identify the Rayleigh coefficients $\left( \alpha_0 , \beta_0 \right)$.}
\label{fig:ratio-3D}
\end{figure}

We now seek to predict the shift of the damping ratios pertaining to modal frequencies that remain in a pre-determined range, throughout the inelastic time history analysis. In other words, we seek an answer to the problem illustrated in figure~\ref{fig:ratio-3D}: How to compute $\Delta > 0$ so that, for $\omega_A$, $\omega_B$, and a targeted damping ratio $\hat{\xi}$ given, for all~$t$ and for all~$m$:
\begin{equation}
 \omega_A \leq \omega_m(t) \leq \omega_B \quad \Rightarrow \quad \hat{\xi} - \Delta < \xi_m(t) < \hat{\xi} + \Delta \quad ?
\end{equation}

We denote $\xi_{max} = \hat{\xi} + \Delta$ and $\xi_{min} = \hat{\xi} - \Delta$. We moreover set $\xi_{max} = \xi_0 (= \xi_A = \xi_B)$ and introduce $R > 1$ so that $\omega_B = R \times \omega_A$. With these notations and considering the expression of the Rayleigh coefficients in equations~\eqref{eq:alp-bet}, we rewrite relation \eqref{eq:xi-abcd} as:
\begin{equation} \label{eq:xi(t)_2}
 \frac{\xi_m(t)}{\xi_{max}} = \frac{1}{R^2 h_B - h_A} \left( R \left( R \ h_B - h_A \right) \frac{1}{ \frac{\omega_m(t)}{\omega_A} } + (R - 1) h_m(t) \frac{\omega_m(t)}{\omega_A} \right) \ .
\end{equation}
To guarantee $\xi_m(t) > 0$, with $h_m(t) > 0$, $\omega_m(t)/\omega_A > 0$, and $R>1$, we require that $h_A>0$ and $h_B>0$ be chosen so that they satisfy the condition:
\begin{equation}
 R \ h_B - h_A \geq 0 \ ,
\end{equation}
which implies that $R^2 h_B - h_A \geq 0$ because $R \geq 1$.

We now study relation~\eqref{eq:xi(t)_2} for methods $a$, $b$ and $d$ explicitly. Because method $b$ is the easiest to deal with, we start with it and methods $a$ and $d$ are considered together right after.

\bigskip

\noindent \textbf{Method $\boldsymbol{b}$:} $h_A = h_B = 1$ and $h_m(t) = 1$ for all $t$ and $m$, leading to the function
\begin{equation} \label{eq:xi(t)_b}
 \frac{\xi_m(t)}{\xi_{max}} = \frac{1}{1 + R} \left( R \frac{1}{ \frac{\omega_m(t)}{\omega_A} } + \frac{\omega_m(t)}{\omega_A} \right) \ ,
\end{equation}
which we write as $\frac{\xi_m(t)}{\xi_{max}} = f \left( \frac{\omega_m(t)}{\omega_A} \right)$. To study the variation of $\xi_m(t)$ with respect to $\omega_m(t)$ in the range $[\omega_A, \omega_B]$, it is useful to compute the derivative $f^{\prime}$ of the function $f$ and to analyze its sign:
\begin{equation}
 f^{\prime} \left( \frac{\omega_m(t)}{\omega_A} \right) = \frac{d \left( \xi_m(t)/\xi_{max} \right)}{d \left( \omega_m(t)/\omega_A \right)} = \frac{1}{1+R} \left( 1 - R \frac{1}{ \big( \frac{\omega_m(t)}{\omega_A} \big)^2 } \right) \ ,
\end{equation}
which implies that $f^{\prime} \left( \sqrt{R} \right) = 0$ and also that $f^{\prime} \left( \frac{\omega_m(t)}{\omega_A} \right) < 0$ for $\omega_A \leq \omega_m(t) < \sqrt{R} \ \omega_A$, and $f^{\prime} \left( \frac{\omega_m(t)}{\omega_A} \right) > 0$ for $\sqrt{R} \ \omega_A < \omega_m(t) \leq \omega_B$, because $R > 1$. Consequently, $\xi_m(t)$ decreases for $\omega_m(t) \in [\omega_A, \sqrt{R} \ \omega_A]$ an then increases for $\omega_m(t) \in [\sqrt{R} \ \omega_A, \omega_B]$ after reaching the minimum:
\begin{equation} \label{eq:xi_min}
 \frac{\xi_m(t)}{\xi_{max}}\bigg|_{min} = \frac{\xi_m(t)}{\xi_{max}}\bigg|_{\frac{\omega_m(t)}{\omega_A} = \sqrt{R}} = \frac{2 \sqrt{R}}{1 + R} \ .
\end{equation}
Because stiffness degradations lead to a decrease of the modal frequencies, the damping ratios in the frequency range $[\omega_A , \sqrt{R} \ \omega_A]$ (respectively $[\sqrt{R} \ \omega_A , \omega_B]$) will necessarily increase (respectively necessarily decrease) during the inelastic analysis. As will be illustrated in the applications that follow this section, this result is useful to design the damping model because it provides information on how to choose the modes and times $A$ and $B$ for which the maximum damping ratio $\xi_{max}$ are observed.

Then, one can compute $\Delta^b$ as follows. By definition, $2 \Delta^b = \xi_{max} - \xi_{min}$, with (see equation~\eqref{eq:xi_min}) $\xi_{min} = \frac{2 \sqrt{R}}{1 + R} \xi_{max}$ and $\xi_{max} = \hat{\xi} + \Delta^b$, which gives:
\begin{equation} \label{eq:Delta}
 \Delta^b = \hat{\xi} \ \frac{1+R-2\sqrt{R}}{1+R+2\sqrt{R}} \ .
\end{equation}
This result is mode- and time-independent. It is a useful measure of the damping ratios drifts in the range $[\omega_A , \omega_B]$, throughout ITHA. Being able to guarantee low $\Delta$ for large $R$ would be an ideal situation to provide good control on the damping ratios. It is however not possible for Rayleigh damping model because $\Delta^b$ monotonically grows with $R$ ($\frac{d \Delta^b}{d R} = \frac{2(R-1)\hat{\xi}}{(1+R+2 \ \sqrt{R})^2} \geq 0$).

An analogous development is presented in~\cite{Hall2006} where the same result is given for $\Delta$ (equation~(5) in~\cite{Hall2006}). But, it should be noted that it is only valid for method $b$, not for methods $a$ or $d$.

\bigskip

\noindent \textbf{Methods $\boldsymbol{a}$ and $\boldsymbol{d}$:} Finding a mode- and time-independent expression for $\Delta$ is in these cases not as straightforward as for method $b$ because the value of $h_m(t)$ is not known \emph{a priori}. First, we define $H$ as a mode- and time-independent variable, such that $h_m(t) \geq H \geq 1$, which, assuming that the stiffness matrix remains positive, is true for all $m$ and $t$ (see equations~\eqref{eq:xi-a} and~\eqref{eq:xi-d}). Then, replacing $h_m(t)$ by $H$, we can rewrite equation~\eqref{eq:xi(t)_2} as:
\begin{equation}
 \frac{\xi_m(t)}{\xi_{max}} \geq \frac{1}{R^2 h_B - h_A} \left( R \left( R \ h_B - h_A \right) \frac{1}{ \frac{\omega_m(t)}{\omega_A} } + (R - 1) H \frac{\omega_m(t)}{\omega_A} \right) = g \left( \frac{\omega_m(t)}{\omega_A} \right)
\end{equation}
and study the variations of the function $g$ with respect to $\omega_m(t)$. From an analogous procedure as for method $b$, we conclude that $g$ decreases for $\omega_m(t) < G \ \omega_A$, is minimum for $\omega_m(t) = G \ \omega_A$, and then increases for $\omega_m(t) > G \ \omega_A$, where $G = \sqrt{ \frac{ R ( R h_B - h_A ) }{ (R-1) H }}$. Consequently:
\begin{equation} \label{eq:xi_min_ad}
 \frac{\xi_m(t)}{\xi_{max}} \geq g(G) = \frac{ 2 \ \sqrt{R (R-1) ( R \ h_B - h_A ) H } }{ R^2 h_B - h_A } \ .
\end{equation}
Because $H \geq 1$, it follows that a mode- and time-independent lower bound for $\xi_{min}$ is:
\begin{equation}
 \xi_{min} \geq \xi_{max} \ \frac{ 2 \ \sqrt{R (R-1) ( R \ h_B - h_A ) } }{ R^2 h_B - h_A } \ .
\end{equation}

In contrast to what was found for method $b$, we only have here an upper bound for $\xi_{min}$. Besides, for method $b$, it is certain that the minimum of $\xi_m(t)$ is attained for $\omega_m(t) \in [\omega_A, \omega_B]$ whereas, for methods $a$ and $d$, we cannot determine \emph{a priori} where the minimum will be observed. Therefore, when designing the damping model, there is in these cases no indication on how to efficiently choose the modes and times $A$ and $B$ for which $\xi_{max}$ is reached.

Considering that $A$ and $B$ have been chosen, and using the same procedure as for method $b$, we obtain the following upper bound for $\Delta^{a,d}$:
\begin{equation} \label{eq:Delta-ad}
 \Delta^{a,d} \leq \Delta^{a,d}_{max} = \hat{\xi} \ \frac{R^2 h_B - h_A - 2 \sqrt{R(R-1)(R h_B - h_A)}}{R^2 h_B - h_A + 2 \sqrt{R(R-1)(R h_B - h_A)}} \ .
\end{equation}
Note that for $h_A = h_B = 1$, we recover $\Delta^{a,d}_{max} = \Delta^b$.

\subsection{Is $\Delta^{a,d}_{max} \leq \Delta^b$?}

\begin{figure}[htb]
\centering
 \includegraphics[width=0.7\textwidth]{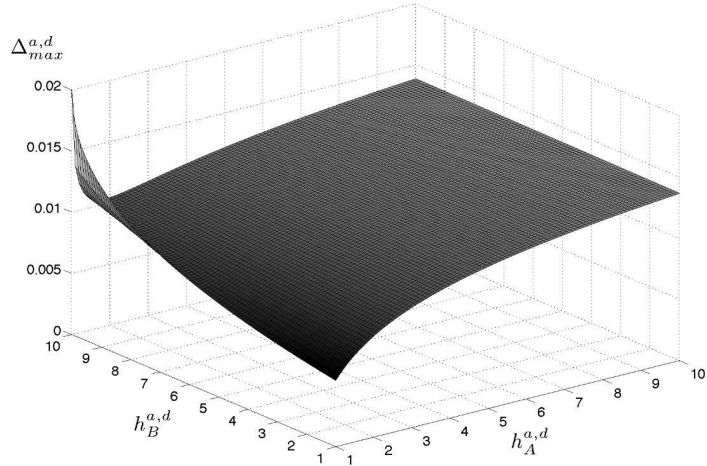}
\caption{$\Delta^{a,d}_{max}$ with respect to $( h^{a,d}_A , h^{a,d}_B )$. $\Delta^{a,d}_{max} = \Delta^b$ for $( h^{a,d}_A , h^{a,d}_B ) = ( 1 , 1 )$. For the purpose of illustration, we use here $R=10$ and $\hat{\xi}=2$\%.}
\label{fig:del-hA-hB}
\end{figure}

If this inequality were true, $\Delta^{a,d}$ would necessarily be lower than $\Delta^b$ for $R$ and $\omega_A$ given and no matter what $h_A$ and $h_B$ would be equal to. Unfortunately, we show in figure~\ref{fig:del-hA-hB} that, for $h_A^{a,d} \geq 1$ and $h_B^{a,d} \geq 1$, we have $\Delta^{a,d}_{max} \geq \Delta^b$, with $\Delta^{a,d}_{max} = \Delta^b$ for $h_A^{a,d} = h_B^{a,d} = 1$. It is thus impossible to conclude anything about whether using initial stiffness-based Rayleigh damping can provide better control on the damping ratios than tangent stiffness-based Rayleigh damping. Besides, one can see in figure~\ref{fig:del-hA-hB} that, in certain area of the $(h^{a,d}_A , h^{a,d}_B )$ plane, $\Delta_{max}^{a,d}$ can be large comparing to $\Delta^b$.

\subsection{Summary}

In table~\ref{tab:sum-cal}, we summarize the formula we derived in this section to calculate the damping ratios time histories and to evaluate their maximum drift. Obviously, Rayleigh damping models designed according to method $b$ have two advantages over models designed with methods $a$ or $d$:
\begin{itemize}
 \item[(i)] The sole quantities that are \emph{a priori} unknown with method $b$ are $\omega_A$ and $R$, which is much easier to approximate with fair accuracy with some experience or simplified analyses than the quantities $h$ with methods $a$ or $d$;
 \item[(ii)] We have access to the exact value of $\Delta^b$, whereas we can only calculate an upper bound for $\Delta^{a,d}$ and it is not straightforward to identify the pair $(t_B , m_B)$ that defines $\xi_B=\xi_{max}$.
\end{itemize}
Nevertheless, we recall that damping models designed from method $b$ can cause solution convergence problems, which is reported in the review of the literature above. It is important to stress here that those models only have the potential to provide practitioners with a better control on the damping ratios time history throughout inelastic seismic analysis. Indeed, as will be illustrated in the next section, it does not imply that a damping model designed from method $a$ or $d$ is necessarily poor.

\begin{table}[htb]
\caption{Initial vs. tangent stiffness-based Rayleigh damping: what the calculus provides. $\xi_{max}=\xi_A=\xi_B$ with $\xi_{A/B} = \xi_{m=m_{A/B}}(t=t_{A/B})$ and $\xi_{max}=\hat{\xi}+\Delta$ with $\hat{\xi}$ the targeted damping ratio in $[\omega_A , R \times \omega_A]$.}
\centering
%\tabsize
\begin{tabular}{lccc}
 \hline
 $i$ & $\xi_m^i(t)/\xi_{max}$ & $\Delta^i/\hat{\xi}$ & Comments \\
 \hline
 \multirow{2}{*}{$a$} & \multirow{2}{*}{$\frac{1}{1 + R} \left( \frac{R}{ \frac{\omega_m(t)}{\omega_A} } + h_m^a(t) \ \frac{\omega_m(t)}{\omega_A} \right)$} & \multirow{2}{*}{$\leq \frac{R^2 h^a_B - h^a_A - 2 \sqrt{R(R-1)(R h^a_B - h^a_A)}}{R^2 h^a_B - h^a_A + 2 \sqrt{R(R-1)(R h^a_B - h^a_A)}}$} & $\cdot$ $h^a_m(t) = \frac{\bphi_m^T(t) \bK_0 \bphi_m(t)}{\bphi_m^T(t) \bK(t) \bphi_m(t)}$ \\
 & & & $\cdot$ $A$ and $B$ \emph{a priori} unknown \\
 $b$ & $\frac{1}{1 + R} \left( \frac{R}{ \frac{\omega_m(t)}{\omega_A} } + \frac{\omega_m(t)}{\omega_A} \right)$ & $=\frac{1+R-2\sqrt{R}}{1+R+2\sqrt{R}}$ & / \\
% $c$ & $\frac{1}{1 + R(t)} \left( \frac{R(t)}{ \frac{\omega_m(t)}{\omega_A} } + \frac{\omega_m(t)}{\omega_A} \right)$ & $\frac{1+R(t)-2\sqrt{R(t)}}{2(1+R(t))}$ & $R$ and $\omega_A$ adjusted through ITHA \\
  \multirow{2}{*}{$d$} &  \multirow{2}{*}{$\frac{1}{1 + R} \left( \frac{R}{ \frac{\omega_m(t)}{\omega_A} } + h_m^d(t) \ \frac{\omega_m(t)}{\omega_A} \right)$} &  \multirow{2}{*}{$\leq \frac{R^2 h^d_B - h^d_A - 2 \sqrt{R(R-1)(R h^d_B - h^d_A)}}{R^2 h^d_B - h^d_A + 2 \sqrt{R(R-1)(R h^d_B - h^d_A)}}$} & $\cdot$ $h^d_m(t) = \frac{\bphi_m^T(t) \bK_0^r \bphi_m(t)}{\bphi_m^T(t) \bK(t) \bphi_m(t)}$ \\
 & & & $\cdot$ $A$ and $B$ \emph{a priori} unknown \\
 \hline
\end{tabular}
\label{tab:sum-cal}
\end{table}

%%%%%%%%%%%%%%%%%%%%%%%%%%%%%%%%%%
\section{Illustrative applications}

\subsection{Example 1: Rayleigh damping designed from elastic structural properties}

The inelastic structural models considered in this example are the same as in Charney's work~\cite{Charney2008}, which we briefly present here. The structure is a five-story building modeled as a system of five DOFs -- the horizontal displacements -- connected by inelastic columns which all have the same elastic properties. At each DOF same mass is lumped. Concerning the inelastic response, two different yielding scenarios are imagined to occur during a hypothetical ITHA:
\begin{itemize}
 \item[(i)] The entire stiffness matrix is assumed to uniformly reduce to 50\% of its original value;
 \item[(ii)] The structural elements nonuniformly yield along the building height: $N$th-story stiffness is reduced to $10\% + (N-1) \times 20\%$ of its original value, that is 10\% for the the 1$^{\tiny \textrm{st}}$ story, 30\% for the 2$^{\tiny \textrm{nd}}$, $\ldots$, 90\% for the 5$^{\tiny \textrm{th}}$.
\end{itemize}

\begin{figure}[htb]
\centering
 \includegraphics[width=\textwidth]{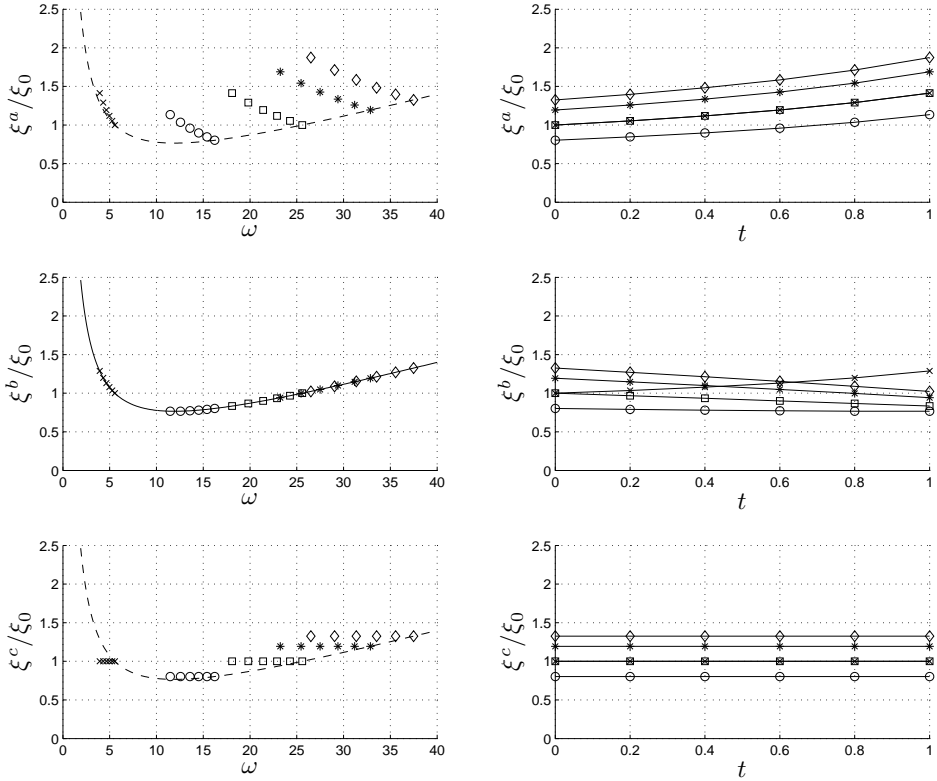}
\caption{Time histories of the damping ratios associated to the five modes of the structure with uniform stiffness degradations (scenario 1). $\times$ is used for mode 1, $\circ$ for mode 2, \tiny $\square$ \small for mode 3, $\star$ for mode 4, and $\diamond$ for mode 5. Time histories are illustrated in $(\xi , \omega)$- and $(\xi , t)$-planes. For method $b$ [center], the curve $\xi^b(\omega)$ is the same for all modes. Conversely, for methods $a$ [top] and $c$ [bottom], $\xi(\omega)$ depends on the mode (the $\times$'s, $\circ$'s,$\ldots$ all belong to a different trajectory in the $(\xi , \omega)$-plane). The curves plotted in the $(\xi , \omega)$-plane are commonly used in the literature to illustrate Rayleigh damping. The dashed lines are plotted at $t=0$.}
\label{fig:char-uni}
\end{figure}

\begin{figure}[htb]
\centering
 \includegraphics[width=\textwidth]{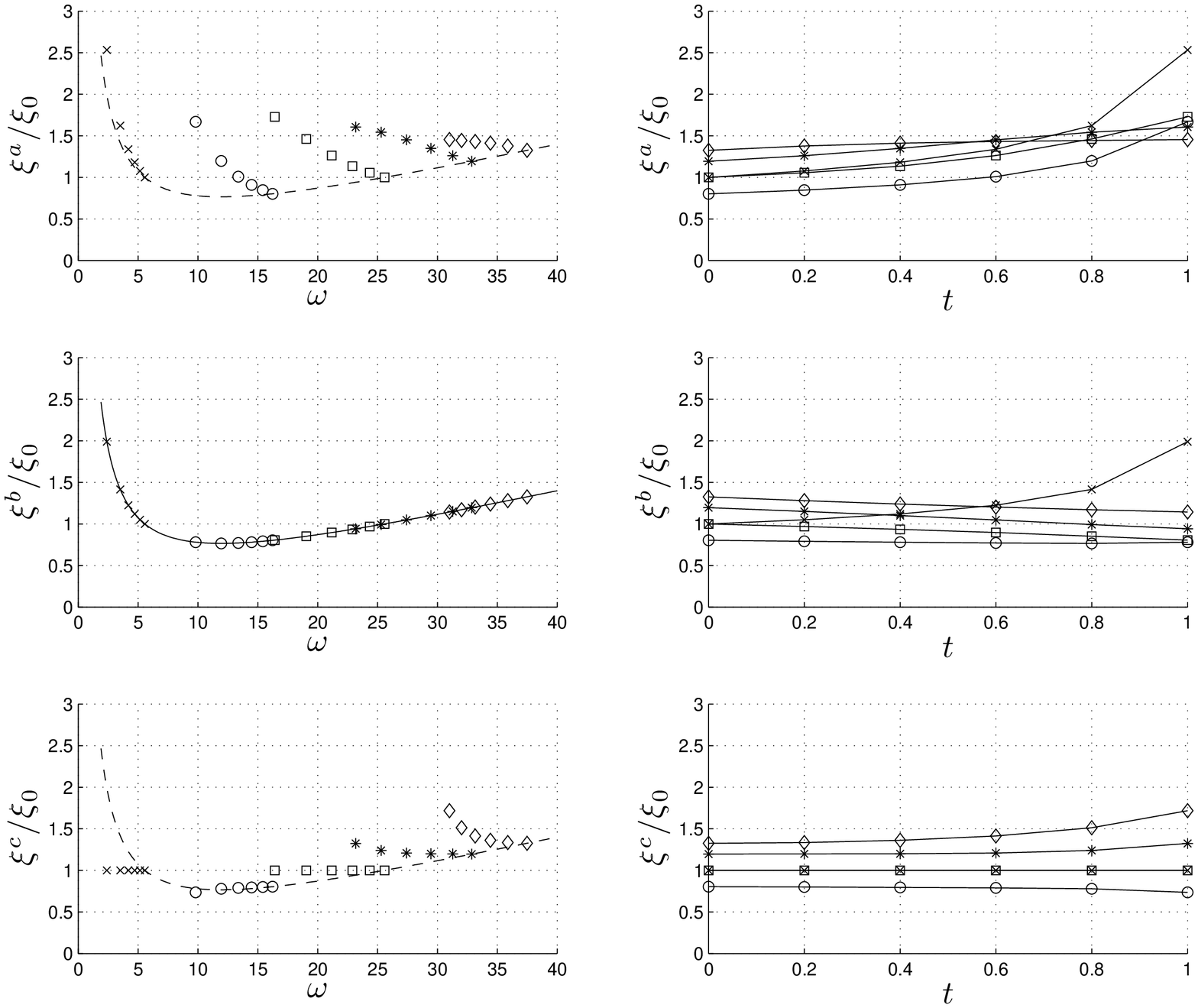}
\caption{Time histories of the damping ratios associated to the five modes of the structure with nonuniform stiffness degradations (scenario 2). Further details provided in the caption of figure~\ref{fig:char-uni} also apply here.}
\label{fig:char-nonuni}
\end{figure}

For this first example, we also design the same Rayleigh damping models as in Charney's work~\cite{Charney2008}, including with method $c$ for a complete comparison. Accordingly, Rayleigh damping models are designed so that a critical damping ratio of 2\% is observed for the modes 1 and 3 of the structure in its initial state ($t = 0$). Adopting the analytical framework presented in the previous section, it means that we design damping models with the following parameters: $\omega_A = \omega_1(0)$, $\omega_B = \omega_3(0)$ and $\xi_0 = \xi_A = \xi_B = \hat{\xi} =$ 2\%. To clearly show the time history of the damping ratios, we arbitrarily set the total duration of the inelastic analysis $T=1$~s and divide the analysis into 5 time steps ($0, 0.2$~s$,$\ldots$, 1$~s). Each time step corresponds to immediate degradations in the structure: \emph{e.g.}, in the case of the uniform yielding scenario described above (scenario 1), at $t=0$ the structural stiffness matrix $\mathbf{K}$ is equal to the initial stiffness matrix $\mathbf{K}_0$, then at $t=0.2$~s the stiffness matrix suddenly changes to $\mathbf{K} = 90\% \times \mathbf{K}_0$, at $t=0.4$~s there is another sudden degradation to $\mathbf{K} = 80\% \times \mathbf{K}_0$, $\ldots$, finally, at $t=1$~s$, \mathbf{K} = 50\% \times \mathbf{K}_0$. The time histories of the damping ratios for methods $a$, $b$ and $c$ are shown in figure~\ref{fig:char-uni} for uniform stiffness degradations (yielding scenario 1) and in figure~\ref{fig:char-nonuni} for nonuniform stiffness degradations (yielding scenario 2).

From figures~\ref{fig:char-uni} and~\ref{fig:char-nonuni}, the best control on the damping ratios time histories is obtained for method $c$, although the damping ratio of mode 5 significantly increases for the structure with nonuniform stiffness degradations. In particular, and as expected for method $c$, modes 1 and 3 are perfectly controlled: in the $(\xi, t)$-plane in figures~\ref{fig:char-uni} [bottom, right] and~\ref{fig:char-nonuni} [bottom, right], $\xi_1(t)$ and $\xi_3(t)$ both describe the same straight line $\xi_1(t) = \xi_3(t) = \xi_0$. We recall that method $c$ is rarely used in practice. It is also obvious that, in this particular case where the Rayleigh damping models are designed according to the initial structural modal properties, method $b$ provides better control on damping than method $a$. For both methods $a$ and $b$, the damping models designed lead to overdamped first modes especially when nonuniform stiffness degradations are assumed. This is not acceptable because first mode generally is important to reconstruct the overall structural behavior. In the case of uniform stiffness degradations, one can also observe in figure~\ref{fig:char-uni} [top, right] that $\xi_1(t)$ and $\xi_3(t)$ describe the same curve and consequently have the same time history.

In table~\ref{tab:ome-h}, we gather the time history of the modal properties of the structure with nonuniform stiffness degradations. In particular, this table shows that the factor $h_m^a$ can become significantly large comparing to their initial unit value.

\begin{table}[htb]
\caption{Time history of the eigenfrequencies [rad.s$^{-1}$] and $h_m^a(t)$ [-] factors for the structure with nonuniform stiffness degradations (see also figure~\ref{fig:char-nonuni}).}
\centering
%\tabsize
\begin{tabular}{c|ccccccccccccccc}
 \hline
 $t$ & & $\omega_1$ & $h_1^a$ & & $\omega_2$ & $h_2^a$ & & $\omega_3$ & $h_3^a$ & & $\omega_4$ & $h_4^a$ & & $\omega_5$ & $h_5^a$ \\
 \hline
 0.0 & & 5.56 & 1.00 & & 16.23 & 1.00 & & 25.58 & 1.00 & & 32.87 & 1.00 & & 37.49 & 1.00 \\
 0.2 & & 5.17 & 1.16 & & 15.42 & 1.11 & & 24.34 & 1.11 & & 31.27 & 1.11 & & 35.87 & 1.09 \\
 0.4 & & 4.72 & 1.41 & & 14.49 & 1.28 & & 22.90 & 1.27 & & 29.45 & 1.26 & & 34.42 & 1.16 \\
 0.6 & & 4.19 & 1.84 & & 13.37 & 1.56 & & 21.18 & 1.54 & & 27.42 & 1.46 & & 33.15 & 1.22 \\
 0.8 & & 3.51 & 2.85 & & 11.94 & 2.13 & & 19.05 & 2.00 & & 25.29 & 1.68 & & 32.02 & 1.27 \\
 1.0 & & 2.39 & 8.10 & &   9.81 & 3.82 & & 16.41 & 2.75 & & 23.18 & 1.89 & & 31.00 & 1.31 \\
 \hline
\end{tabular}
\label{tab:ome-h}
\end{table}

\subsection{Example 2: Design of optimal Rayleigh damping models}

In this example, the same structure as for example 1 is considered, but only with nonuniform stiffness degradations (yielding scenario 2), which is the more realistic case.

When choosing $\omega_A$, $\omega_B$, $\xi_A$ and $\xi_B$, the practitioner has to seek for the best control on all the damping ratios pertaining to the most important modes, throughout the ITHA. When structure yields, the modal frequencies decrease, which can lead to overdamping modes, especially for the fundamental one which can increase a lot from its initial value to its actual value after yielding (see figure~\ref{fig:char-nonuni} where it increases by a factor of up to $2.5$). In this second example, we present how to design optimal Rayleigh damping models with methods $a$ (or $d$) and $b$ to provide the best control possible on the damping ratios time histories.

To that purpose, and according to the analytical developments presented in the previous section, we now design Rayleigh damping models accounting for stiffness degradations:
\begin{itemize}
 \item[(i)] As above, we consider that assigning a damping ratio to modes 1 and 3 is relevant to control the damping ratio of the most important modes and thus set $m_A=1$ and $m_B=3$. We remark that $\omega_1$ and $\omega_3$ necessarily remain in the range $[\omega_1(T) , \omega_3(0)]$ throughout ITHA because $\omega_1(T) < \omega_1(0)$ and $\omega_3(0) > \omega_3(T)$ due to stiffness degradations, where $T$ is the total duration of the simulation. Note that the quantities at time $T$ are \emph{a priori} not known and need some preliminary analysis results to be efficiently chosen; hereafter, we will use the results previously obtained with example 1. Then, the choice of $t_A$ and $t_B$ depends on the method used:
\begin{itemize}
 \item For method $a$: according to figure~\ref{fig:char-nonuni}, $\xi^a_1(t)$ and $\xi^a_3(t)$ constantly increase throughout the analysis. Consequently, the maximum damping ratios for modes 1 and 3 will be observed at time $t = T$ and this is why we set $t_A=t_B=T$. Then, $\omega_A = \omega_1(T) = 2.39$~rad.s$^{-1}$, $\omega_B = \omega_3(T) = 16.41$~rad.s$^{-1}$ ($R=6.87$), $h_A = h_1(T) = 8.10$, and $h_B = h_3(T) = 2.75$;
 \item For method $b$: we set $t_A=T$ and $t_B=0$, because, according to figure~\ref{fig:char-nonuni}, $\xi_{1, max}=\xi_1(T)$ and $\xi_{3, max}=\xi_3(0)$. Then, $\omega_A = \omega_1(T) = 2.39$~rad.s$^{-1}$ and $\omega_B = \omega_3(0) = 25.58$~rad.s$^{-1}$ ($R=10.70$).
\end{itemize}
 \item[(ii)] With a targeted damping ratio in the range $[\omega_A , R \ \omega_A]$ ($R \ \omega_A = \omega_B$) of $\hat{\xi}=2\%$, we compute:
\begin{itemize}
 \item For method $a$: $\Delta^a_{max} = 1.06$\% from equation~\eqref{eq:Delta-ad}. $\Delta^a_{max}$ is an upper bound and a lower value is therefore likely to be more appropriate. Nevertheless, we take $\xi_{max} = \xi_0 = \xi_A = \xi_B = 2\% + 1.06\% = 3.06\%$, which has to be validated once the analysis has been run;
 \item For method $b$: $\Delta^b=0.57$\% from equations~\eqref{eq:Delta} and we set $\xi_{max} = \xi_0 = \xi_A = \xi_B = 2\%+0.57\%=2.57$\%.
\end{itemize}
\end{itemize}

\begin{figure}[htb]
\centering
 \includegraphics[width=1\textwidth]{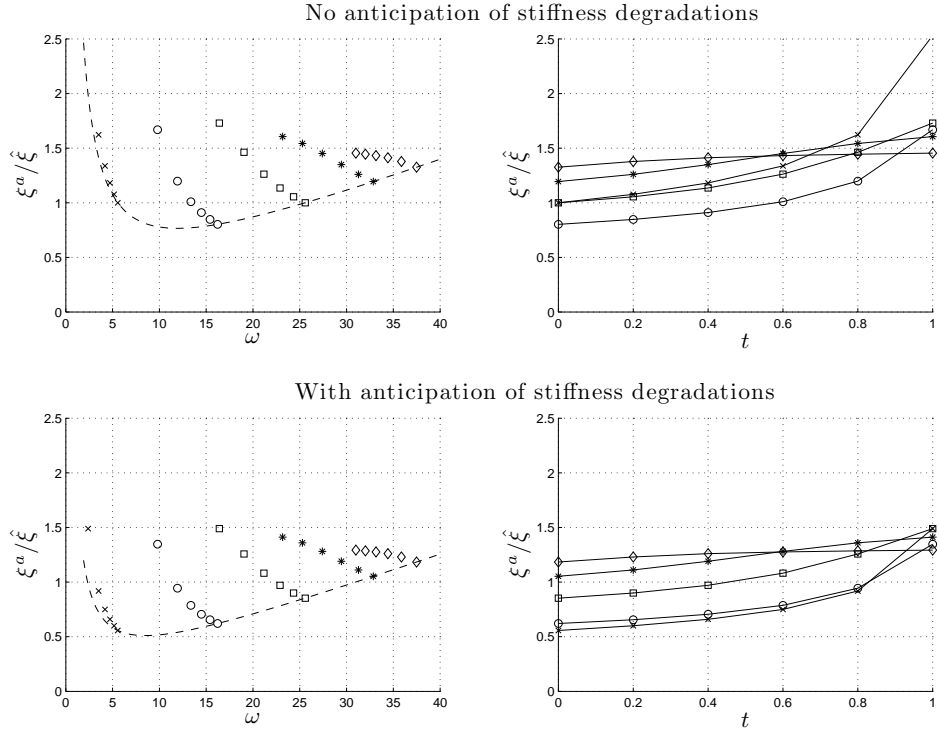}
\caption{Structure with nonuniform stiffness degradations. [bottom] Rayleigh damping model of type $a$ with anticipation of stiffness degradations. [top] For the sake of comparison, the figures reproduce the results shown in figure~\ref{fig:char-nonuni} where stiffness degradations are not accounted for in the design of the damping model.}
\label{fig:imp-xi-a}
\end{figure}

\begin{figure}[htb]
\centering
 \includegraphics[width=1\textwidth]{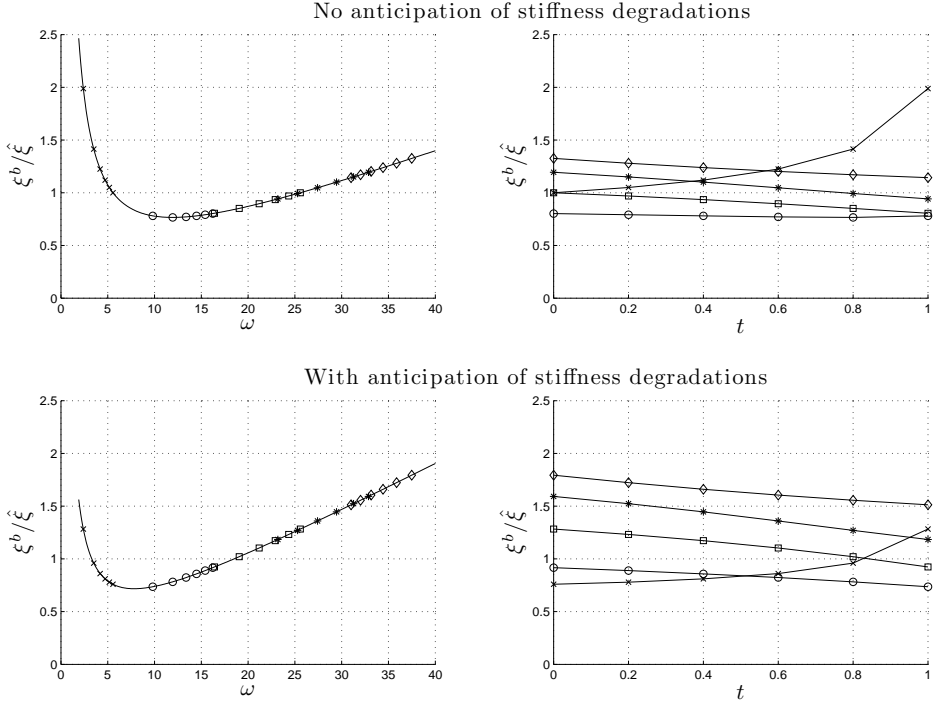}
\caption{Structure with nonuniform stiffness degradations. [bottom] Rayleigh damping model of type $b$ with anticipation of stiffness degradations. [top] For the sake of comparison, the figures reproduce the results shown in figure~\ref{fig:char-nonuni} where stiffness degradations are not accounted for in the design of the damping model.}
\label{fig:imp-xi-b}
\end{figure}

The histories of the damping ratios resulting from the Rayleigh damping models designed according to the procedure presented just above are shown in figure~\ref{fig:imp-xi-a} when method $a$ is used and in figure~\ref{fig:imp-xi-b} for method $b$. For the sake of comparison, the damping ratios time histories shown in figure~\ref{fig:char-nonuni}, \emph{viz.} when stiffness degradations are not accounted for in the damping models design, are reproduced in the top half of figures~\ref{fig:imp-xi-a} and~\ref{fig:imp-xi-b}.

According to figure~\ref{fig:imp-xi-a}, one can make the following observations for method $a$:
\begin{itemize}
 \item[(i)] The damping ratios pertaining to the first three modes remain in the range $[1.11\% , 2.98\%]$ which is, as expected after the analytical developments in the previous section, included in the range $[0.94\% , 3.06\%] = [\hat{\xi} - \Delta^a_{max} , \hat{\xi} + \Delta^a_{max}]$. The choice of $\xi^a_{max} = \xi^a_1(T) = \xi^a_3(T)$ thus is validated.
 \item[(ii)] $\hat{\xi}=2\%$ is well centered in $[1.11\% , 2.98\%]$, which means that the choice of $\xi^a_{max} = \xi^a_A = \xi^a_B = \hat{\xi} + \Delta^a_{max}$ was satisfying.
 \item[(iii)] Control on the damping ratios is significantly improved when stiffness degradations are anticipated, especially for the first mode. 
 \item[(iv)] Although modes 4 and 5 are out of the selected control range $[\omega_1(T) , \omega_3(T)]$, they are less overdamped than when the damping model does not account for stiffness degradations.
\end{itemize}

Looking now at figure~\ref{fig:imp-xi-b}, we can remark the following for method $b$:
\begin{itemize}
 \item[(i)] The damping ratios pertaining to the first three modes remain in the range $[1.47\% , 2.57\%]$ which is, as expected, included in the range $[1.43\% , 2.57\%] = [\hat{\xi} - \Delta^b , \hat{\xi} + \Delta^b]$. We could decrease $\xi^b_{max}$ of $0.02\%$ to have $\hat{\xi}$ perfectly centered in $[1.47\%-0.02\% , 2.57\%-0.02\%] = [1.45\% , 2.55\%]$: the correction is minor in this example because the minimum of the $\xi^b(\omega)$ curve is almost reached by $\xi^b_3(T)$ (see figure~\ref{fig:imp-xi-b} [bottom])\footnote{Such a simple \emph{a posteriori} correction of $\xi_{max}$ is possible here because $h_A=h_B=1$ and $\xi_A=\xi_B$.}.
 \item[(ii)] Control on the first damping ratio is significantly improved when stiffness degradations are anticipated. 
 \item[(iii)] Higher modes are overdamped. We could increase $R$ to have a better control on these modes too, but this would increase $\Delta^b$ and consequently alter the control on the first three modes.
\end{itemize}

Finally, comparing figure~\ref{fig:imp-xi-a} [bottom] to~\ref{fig:imp-xi-b} [bottom], we remark that:
\begin{itemize}
 \item[(i)] A Rayleigh damping model designed with method $a$ with anticipation of stiffness degradations can be more efficient than designed with method $b$ without accounting for stiffness degradations. In particular, it avoids the strong increase of the damping ratio pertaining to the first mode.
 \item[(ii)] As far as the first three modes are concerned, the control provided by method $b$ is better: $[1.43\% , 2.57\%] \subset [1.11\% , 2.98\%]$.
 \item[(iii)] Better control on the modes 4 and 5 is observed with method $a$. However, Rayleigh damping models were not designed to control those two modes. For method $b$, one could have better control on modes 4 and 5, by increasing $\omega_B$, and so $R$, until $\Delta^b$ reaches the actual $\Delta^a=0.98\%$.
\end{itemize}

\subsection{Some additional guidelines for practical use}

As illustrated above, it is necessary to have some hint of the overall nonlinear structural behavior to design a Rayleigh damping model that efficiently controls the damping ratio time history for the structural modes of interest. This issue is inherent to every Rayleigh damping model in ITHA. In practice, the time history of the modal properties are however not known \emph{a priori} and designing optimal Rayleigh damping model is thus not straightforward, especially with method $a$, and might need some iterations. The following practical guidelines can therefore be useful:
\begin{itemize}
 \item[(i)] It is recommended in \cite{LegDus1992} that $\omega_A$ be the first mode and $\omega_B$ the lowest mode for which the cumulative effective mass exceeds 90\%-95\% of the total mass.
 \item[(ii)] The following possible methods are recommended in the user manual of the computer program PERFORM-3D~\cite[\S 18.2]{Perform3D} to define $\omega_A$ and $\omega_B$:
\begin{equation} \label{eq:gamma}
 (i) \ \left\{
 \begin{array}{l}
  \omega_A = 1.10 \ \omega_1(0) \\
  \omega_B = 4.00 \ \omega_1(0)
 \end{array}
 \right.
 \qquad \textrm{or} \qquad
 (ii) \ \left\{
 \begin{array}{l}
  \omega_A = 1.10 \ \omega_1(0) / \sqrt{\mu} \\
  \omega_B = 0.85 \ \omega_1(0)
 \end{array}
\right. \ ,
\end{equation}
where $\mu$ is the ductility of the structure. Method (i) is the same as for a linear analysis and method (ii) is particularly effective for structures dominated by their first mode.
\end{itemize}

%%%%%%%%%%%%%%%%%%%%%%%%%%%%%%%%%%
\section{Conclusions}

From the review of the literature proposed in the second section of this paper, it is obvious that some researchers or practitioners advocate using the initial stiffness matrix in the design of Rayleigh damping models, whereas others advocate using the tangent stiffness matrix. Controlling the damping ratios throughout inelastic time history analyses is an important issue to avoid the appearance of spurious damping forces. To that purpose, useful analytical formulas are developed in the third section of this paper for both Rayleigh models based on initial and tangent stiffness. Whereas there exists a simple relation that allows controlling the damping ratios time histories when tangent stiffness is used, there is no such relation when initial stiffness is used and controlling damping ratios is, although achievable, not straightforward.

From these latter analytical relations and from the examples shown in section 4, we can conclude that it is easier to design a Rayleigh damping model with well-controlled damping ratios time histories throughout the inelastic analysis when the tangent stiffness is used than with the initial stiffness. However, controlling the damping ratios with initial stiffness-based damping models can be achieved. That is why, when convergence difficulties are experienced with the solution algorithm, initial stiffness can be used without necessarily leading to very high damping ratios.

Whether initial or tangent stiffness is used in the design of the Rayleigh damping model used for inelastic time history seismic analysis, we advocate that:
\begin{itemize}
 \item[(i)] Figures like figures~\ref{fig:imp-xi-a} and~\ref{fig:imp-xi-b} in this paper be plotted to keep control on the damping ratios time histories. To that purpose, the required analytical relations are summarized in table~\ref{tab:sum-cal} of this paper. These later relations also help improving control on the damping ratios;
 \item[(ii)] The damping forces be computed and compared to the other resisting forces so as to guarantee there are no spurious damping forces generated in the system. Indeed, even if the damping ratios are well-controlled throughout the inelastic time history seismic analysis, Rayleigh damping still lacks physical evidence and there is no guarantee that the actual damping forces are properly modeled.
\end{itemize}

Considering the difficulty to control Rayleigh damping in inelastic structures that is illustrated in this paper, strategies that rely on using models with nonlinearities expected to develop in clearly identified structural parts with the other parts remaining elastic appear as promising. Such strategies are mentioned in the second section of this paper but are out of the scope of the developments that follow in sections 3 and 4. However, there are some obvious arguments that would motivate further investigations for this recent class of methods in future work: i) in elastic parts, there is no concerns about damping ratio shift; ii) a reduced -- in comparison to the number of structural DOFs -- eigenbasis could be defined (\emph{e.g.} with the procedures proposed in~\cite{IbrWil1992} and~\cite{IbrWil1989}), which would increase computational efficiency; iii) this would simplify the damping model design because the number of modes, which a proper damping ratio has to be associated to, would be reduced; iv) the inelastic parts could be separately treated with practical methods specifically developed to avoid the problems likely to be encountered when using Rayleigh damping.

%%%%%%%%%%%%%%%%%%%%%%%%%%%%%%%%%%%%%
\vspace{0.7cm}

\textbf{Acknowledgements} This research was supported by a Marie Curie International Outgoing Fellowship within the 7th European Community Framework Programme (proposal No. 275928). The financial support provided by the Fond Qu\'eb\'ecois de Recherche sur la Nature et les Technologies (FQRNT), the Natural Science and Engineering Research Council of Canada (NSERC), and the ENS-Cachan Invited Professor Program also are gratefully acknowledged.

%%%%%%%%%%%%%%%%%%%%%%%%%%%%%%%%%%%%%

\bibliographystyle{plain}
\bibliography{EESD_2013_PJ-PL-AI}

\end{document}